\documentclass[preprint, times]{elsarticle}

\usepackage{amssymb}
\usepackage{amsthm}
\usepackage{mathtools}

\usepackage{textcomp}

\usepackage{siunitx}

\usepackage{hyperref}

\usepackage{textgreek}

\usepackage{dblfloatfix}
\usepackage{fixltx2e}
\usepackage{float}
\usepackage{booktabs}
\usepackage{tabulary}

\usepackage{comment}

\DeclareGraphicsExtensions{.pdf, .eps, .jpg}
\journal{NIMA}

\begin{document}

\begin{frontmatter}

\title{Design of a large dynamic range readout unit for the PSD detector of DAMPE}

\author[imp,lzu,ucas]{Yong Zhou}
\author[imp]{Zhiyu Sun}

\author[imp]{Yuhong Yu\corref{corresponding_author}}
\cortext[corresponding_author]{Corresponding author}
\ead{yuyuhong@impcas.ac.cn}

\author[imp]{Yongjie Zhang}
\author[imp]{Fang Fang}
\author[imp]{Junling Chen}

\author[lzu]{Bitao Hu}

\address[imp]{Institute of Modern Physics, Chinese Academy of Sciences,  509 Nanchang Road,  Lanzhou,  730000,  P.R.China}
\address[lzu]{School of Nuclear Science and Technology,  Lanzhou University,  222 South Tianshui Road,  Lanzhou,  730000,  P.R.China}
\address[ucas]{Graduate University of the Chinese Academy of Sciences,  19A Yuquan Road,  Beijing,  100049,  P.R.China}
\begin{abstract}

A large dynamic range is required by the Plastic Scintillator Detector (PSD) of DArk Matter Paricle Explorer (DAMPE), and a double-dynode readout has been developed. To verify this design, a prototype detector module has been constructed and tested with cosmic rays and heavy ion beams.
The results match with the estimation and the readout unit could easily cover the required dynamic range.

\end{abstract}


\end{frontmatter}

\section{Introduction}
\label{sec:introduction}

DArk Matter Paricle Explorer (DAMPE) is a satellite-borne particle
detector aiming for in-direct dark matter search, high-energy gamma
astronomy and primary cosmic ray study~\cite{Chang_Jin_dampe}. It is
designed to cover a wide energy range, from \SI{5}{GeV} to
\SI{10}{TeV} for electrons and photons and from \SI{10}{GeV} to
\SI{1}{PeV} for heavy ions, with unprecedented
resolution(\SI{1.5}{\percent} for \SI{100}{\giga\electronvolt} electrons).

The Plastic Scintillator Detector (PSD), located at the top of the
satellite, is a key component of DAMPE. It consists of two layers
of plastic scintillator bars which are orthogonal to each other and
covers an effective area of
$\SI{820}{mm}\times\SI{820}{mm}$. The bars, with a dimension of
$\SI{884}{mm} \times \SI{28}{\milli\meter} \times \SI{10}{mm}$, are
made of EJ-200~\cite{scintillator}, and readout at each end by a
Hamamatsu R4443 photomultiplier tub (PMT) , which is a
low noise PMT ruggedized for space usage~\cite{r4443}. The EJ-560 optical pads are used for
coupling the bars and the PMTs, which will assure the proper transmission of the scintillation lights.

By measuring the deposited energy, the PSD serves as an
anti-coincidence detector for e/$\gamma$ discrimination as well as a
charge detector for heavy ions up to Z=20, and these requires the readout
system of PSD to cover a wide range of signal amplitudes. On the other hand,
the limitations on available weight and power imposed by the satellite is also an
extra constraint in the design of the readout unit.

In this paper, a large dynamic range readout unit with low power consumption for PSD is reported.
Verification tests for the design have been carried out using cosmic ray and relativistic heavy ion beam, and the results are satisfactory.
The detailed description of the design is presented in Sec.\ref{sec:requirement} and Sec.\ref{sec:design}, and the test results are discussed in Sec.\ref{sec:result}.

\section{Dynamic range requirement}
\label{sec:requirement}

The dynamic range of the detector maybe influenced by many different factors. To simplify the work, we define the mean light yield produced by a minimum ionizing particle (electron with $\gamma$ = 5 for example) penetrating vertically through the center of the bar as one MIP, and use it as the unit for the dynamic range estimation.

The relationship between the response of the scintillator to a charged particle and its scintillation efficiency is one of the most important aspects, which will enlarge the dynamic range requirments. The light yield of organic scintillator is a nonlinear function of the specific energy loss $dE/dx$ due to the quenching effect.
It can be described precisely by the Birks' law~\cite{birks_theory_2013} in the low energy range, while for heavy ions in the relativistic region, the scintillation mechanism is much more complicated and the Birks' law fails to apply directly. Several modified models~\cite{chou_nature_1952,tarle_cosmic_1979,menchaca-rocha_response_1999,matsufuji_response_1999} have been proposed to extend the Birks law to the high energy region, nevertheless large discrepancy exists between them when extrapolating to the high-Z nuclides.

Due to the theoretical difficulties mentioned above, our estimation of the light yield is based on the direct relativistic heavy ion beam test result of AMS-02 TOF prototype~\cite{bindi2005performance}, which also uses EJ-200 as the detection material and has the same thickness.
According to the parameters of~\cite{bindi2005performance}, the light output of calcium (Z=20) is about 270 times larger than that of proton (Z=1).
As all minimum ionization particles with Z=1 have approximately the same specific energy loss in the organic material, it means that the light output for all the high energy particle species detected by PSD has a maximum of 270 times difference.

The second factor we considered is the broad field view of DAMPE, which asks for a maximum incidence angle of \SI{60}{\degree}. It indicates that the traversing length and the energy deposit in the bar is double to the perpendicular incident particle, and this expands the maximum light yield to \SI{540}{MIPs}. Moreover, the statistical nature of the ionizing process should be also taken into account. Assuming the energy resolution of the fluctuation is \SI{10}{\percent} ($\sigma$), an upper limit of the measurable energy deposition per scintillator bar up to 675 MIPs is required as a 5$\sigma$ width is accepted. And the lower limit of measurement must be smaller than \SI{0.75}{MIP} on the other hand.

The third factor is the special shape of the scintillator bar since the light attenuation along the bar could not be negligible in the transmission process.
To improve reflection efficiency, the bars are surface-polished and wrapped by the Tyvek paper~\cite{tyvek}. A preliminary test shows that a reduction ratio in the intensity of the scintillation light can be controlled to \SI{50}{\percent} as the incident particle hit the center position perpendicularly. Thus, the light collected by the PMTs at two ends will vary between \SI{0.375}{MIP} and \SI{1350}{MIPs}.

Considered the noise from PMTs and electronics and to ensure the identification of even the smallest signal from noise clearly and leave some redundancy for the adjustment, the final dynamic range demand for the readout unit of PSD is determined to be \SI{0.1}{MIP}$\sim$\SI{1400}{MIPs}.

\section{Design of the readout unit}
\label{sec:design}

\subsection{Readout scheme}
\label{sec:scheme}

\begin{figure*}
\centering
 \includegraphics[width=140mm]{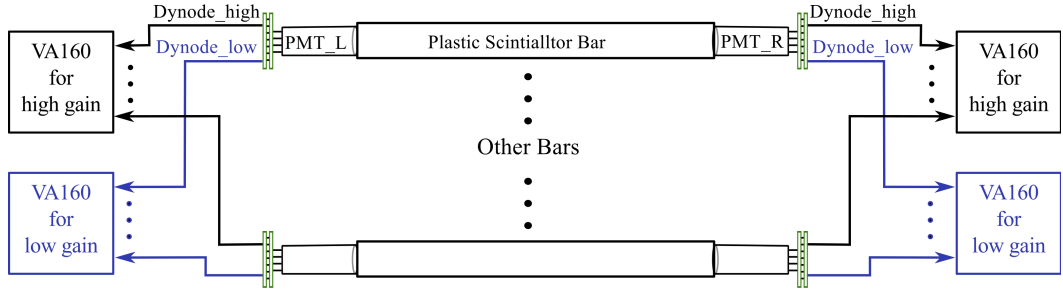}
\caption{PSD readout scheme with two measurement ranges.}
\label{fig:readout_scheme}
\end{figure*}

 An ASIC chip VA160, which is developed by IDEAS~\cite{va160} and optimized for the DAMPE project, is adopted as the major signal processing circuit for the charge measurement of PSD.
 The VA160, with a dynamic range from \SI{-3}{\pico\coulomb} to \SI{13}{\pico\coulomb}, is a low power consumption and radiation harden chip, and optimized for positive input signals from the PMT dynodes output.

To measure the energy deposition in the range with 4 orders of magnitude at each scintillator bar, it's found that a single VA160 channel is not sufficient to cover the full dynamic range required by the PSD readout unit. A readout system that uses a PMT with double dynode outputs coupled to separate VA160 channels has been developed and performed by the PSD group. Two measurement ranges can be achieved in this way, with the high-gain dynode channel for small light output measurement and the low-gain dynode channel for large light output measurement. To reduce the crosstalk between low-gain and high-gain channels, signals from the two dynodes are connected to different VA160 chips as shown in Fig.\ref{fig:readout_scheme}.

The FEE board of PSD which is based on VA160 has been developed and fully tested~\cite{fee}. The analog to digital conversion is performed by a 14-bit ADC. A linear range up to \SI{12}{\pico\coulomb} is guaranteed for each FEE channel. An input noise level of the whole PSD is expected to not greater than \SI{6}{\femto\coulomb}. This implies the response of \SI{0.1}{MIPs} shall be larger than \SI{30}{\femto\coulomb} as a 5$\sigma$ separation is required, i.e. 1 MIP$\geq$\SI{300}{\femto\coulomb}.
 Thus a dynamic range from \SI{0.1}{MIPs}$\sim$\SI{40}{MIPs} can be achieved by the high-gain dynode channel, and the remaining range shall be covered by the low-gain dynode channel.
This constrains the gain ratio between the two dynodes should be larger than $\geq$35.

\subsection{Selection of the dynode stages}
\label{sec:dynodes_selection}
To determine the appropriate dynode stages for readout, an estimation of the mean number of photon electrons (PE) generated by one MIP in the center of the bar is carried out.

The mean energy deposit of the MIP particle in a \SI{10}{\milli\meter} plastic scintillator is about \SI{2.0}{\mega\electronvolt}~\cite{olive_review_2014}, and the scintillation efficiency of EJ-200 is \SI[per-mode=symbol]{e4}{photons\per\mega\electronvolt}~\cite{scintillator}.
Light is distributed to both ends of the bar equally.
However, only \SI{22.5}{\percent} of the generated photons are within the total reflection angle and will undergo the transmission process to the end of the bar, due to the refraction on the boundary surface.
The final number of photon electrons is estimated as follows:
\begin{align}
 N_{PEs} &= \frac{1}{2} \times \SI[per-mode=symbol]{2}{\mega\electronvolt} \times \SI{e4}{\per\mega\electronvolt} \times 0.225
           \times \varepsilon_{1} \times \varepsilon_{2} \times \varepsilon_{3} \times \varepsilon_{4} \nonumber \\
         &\approx \SI{48}{PEs\per{MIP}}
\label{eq:pes}
\end{align}
where $\varepsilon_1$ ($\approx$0.5) is the light attenuation ratio from the center to the end of the bar,
$\varepsilon_2$ ($\approx$0.3) is the the geometry factor induced by the effective coupling area between the PMT and the end surface of the scintillator bar,
$\varepsilon_3$ ($\approx$0.95) is the transmission coefficient of the silicon rubber which is the coupling material between the readout device and the bar,
and $\varepsilon_4$ ($\approx$0.15) is the mean quantum efficiency of the PMT.

The PSD will use R4443, which is a 10-stage PMT, as the readout device. By using a standard voltage divider (i.e. equal voltage drop between dynodes), the gain of R4443 is $1\times10^6$ at the typical supply voltage, which is obviously too high for the VA160 chip. A simple calculation shows that the secondary emission ratio $\delta$ between adjacent dynodes is about 3.98 as the standard voltage is applied according to Eq.~\ref{eq:gain}.
\begin{equation}
G=\delta^{n}
\label{eq:gain}
\end{equation}
Where $G$ is the anode current amplification, and $n$ is the number of the dynodes. To match the dynamic range of VA160, the signal from the middle dynodes is extracted to reduce the gain factor of the PMT, and the n-th dynode signal charge is given by
\begin{equation}
Q_{dn}=Q_{n-1}-Q_{n}
\label{eq:dycharge}
\end{equation}
Where $Q_{n-1}$ is collected charge of the n-th dynode, $Q_{n}$ is emitted charge of the n-th dynode. Since the secondary emission ratio $\delta$ has the same value per stage for the given voltage distribution, the gain of the n-th dynode is given by Eq.~\ref{eq:dygain}.
\begin{equation}
G_{dn}=\delta^{n}(1-1/\delta)
\label{eq:dygain}
\end{equation}
The 8th dynode (Dy8), with a gain of $4.71\times10^4$ from Eq.~\ref{eq:dygain}, has a positive charge output \SI{362} {\femto\coulomb} for one MIP signal. It's a little larger than our expectation 300 fC, and the difference can be eliminated easily by fine adjustment of the PMT supply high voltage. To cover the required upper dynamic range, an earlier stage 5th dynode (Dy5) is used to the larger signals measurements, as the maximum relative gain factor across three dynodes is \SI{63} from a numerical calculation. Therefore, the total readout system could achieve the demanded wide range of 4 orders, and also provide a cross calibration between this two dynodes.

\subsection{Voltage divider circuit}
\label{sec:divider_design}

\begin{figure*}
\centering
 \includegraphics[width=130mm]{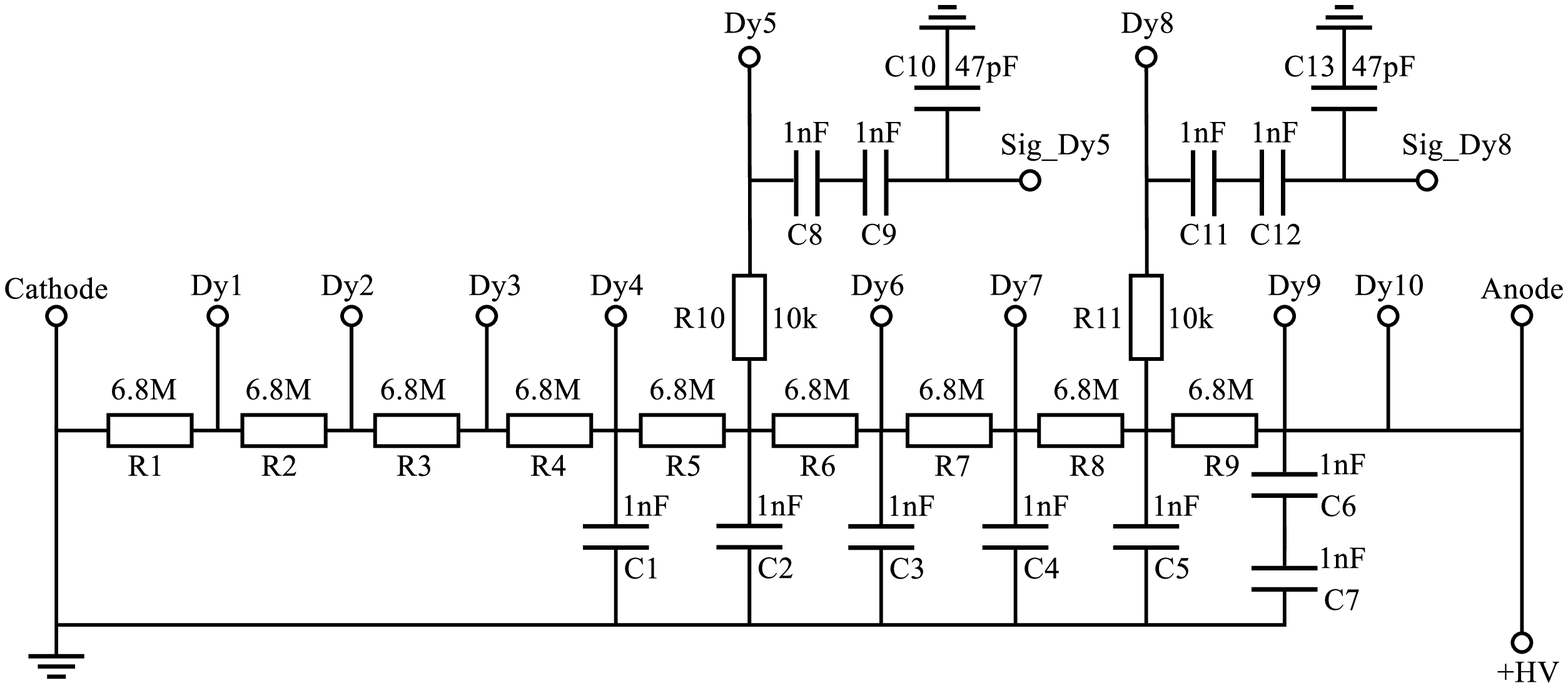}
\caption{Circuit diagram of the voltage divider.}
\label{fig:divider}
\end{figure*}

A cathode grouding voltage divider circuit, extracted charge pulses from double dynodes Dy5 and Dy8, is designed as shown in Fig.\ref{fig:divider}.
The uniform voltage distribution ratio is adopted instead of the tapered one because the space charge effects is not an issue at the PSD light intensity level.

In this design, the parallel decoupling capacitors C1$\sim$C7 are added to keep the last few interstage voltage stable. Two serial-chained capacitors (C8/C9 and C11/C12) are used in each branch to provide robustness, in case that one of them breaks down and the other can still protect the following circuits. The damp resistors R10/R11 and the bypass capacitors C10/C13 are added to better form the output pulse shape. To minimize the power consumption of the PSD, the power of each PMT divider is designed to less than 20 mW. Moreover,the coaxial cables are used to transmit  the signals from different dynodes but the same PMT to the FEE board to reduce the potential for signal interference.

\section{Verification tests}
\label{sec:result}
Based on the design described above, a prototype of the PSD detector module has been developed and some tests with cosmic ray and high energy ion beams are carried out to check the performance of the readout unit.

\subsection{Results from cosmic ray tests}
\label{sec:cosmicray}

\begin{figure*}
 \centering
 \includegraphics[width=130mm]{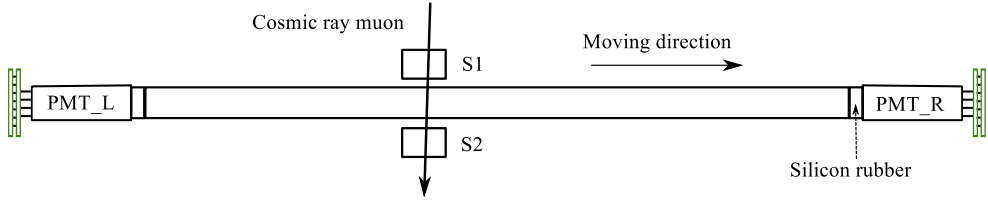}
\caption{The experimental setup for the cosmic ray test.}
\label{fig:cosmic_test}
\end{figure*}

Cosmic ray muons are the most numerous energetic charged particles at the sea level, which are also minimum ionization particles. So they are broadly used to study the MIP response of detectors at the laboratory.
We have used the cosmic ray muons to test the PSD prototype also and the layout of the test setup is shown in Fig.~\ref{fig:cosmic_test}.
Two small plastic scintillator detectors ($\SI{10}{mm} \times \SI{10}{mm} \times \SI{10}{mm}$) are used to tag the vertically incident cosmic ray, and they are fixed in a movable support to investigate the response difference at various positions along the bar.

\begin{figure}
    \centering
    \includegraphics[width=90mm]{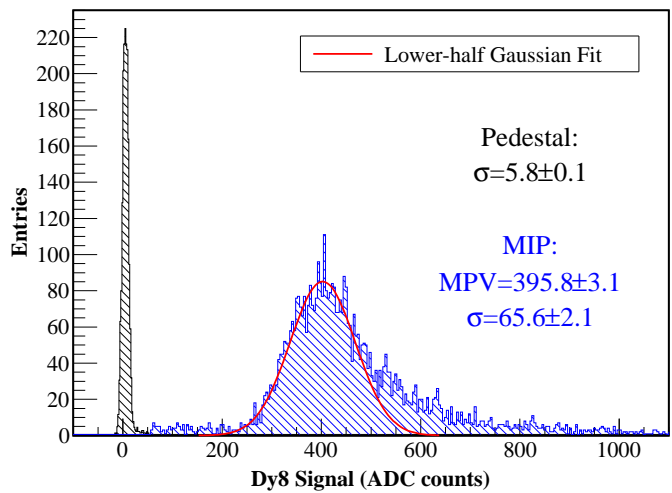}
    \caption{The MIPs response recorded by dynode 8 in the middle of the PSD scintillator bar. The noise spectrum of this readout channel are shown in the same figure.}
    \label{fig:mip}
\end{figure}

The result from the right side of the prototype module is shown in Fig.\ref{fig:mip}.
The spectrum follows the Landau distribution and the most probable value (MPV) is 396 ADC counts (\SI{366}{\femto\coulomb}).
By fitting the lower half of the MIPs peak using Gaussian distribution, the relative resolution is found to be about \SI{16.6}{\percent}.
The pedestal spectrum of this readout channel is also shown in this figure, and the result of the linearity test from the FEE board is shown in Table ~\ref{tab:summary}. The value of the RMS noise is 5.8 ADC counts (\SI{5.4}{\femto\coulomb}), which implies the lower limit of the dynamic range for this readout unit is about \SI{0.08}{MIPs} as a 5$\sigma$ separation is adopted to distinguish signal to noise.

\begin{figure}
    \centering
    \includegraphics[width=90mm]{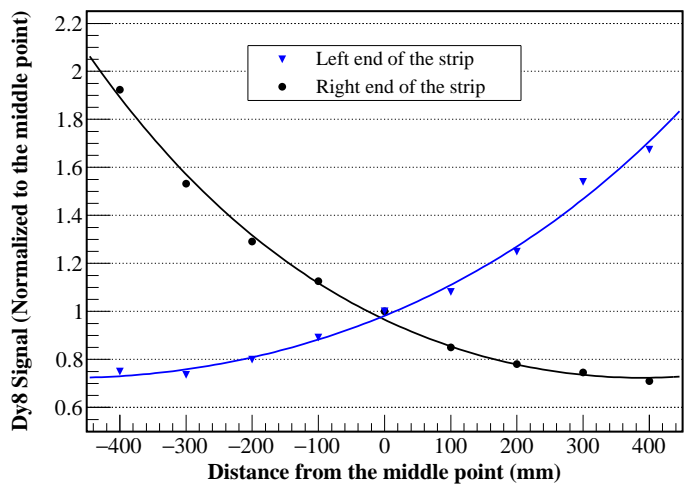}
    \caption{The relative light output measured by the end PMTs at different hit positions along the PSD scintillator bar.}
    \label{fig:attenuation}
\end{figure}

The MIPs measurements are then carried out every \SI{10}{\centi\meter} along the bar to investigate the influence of the light attenuation in detail.
The relative light output at each position is shown in Fig.~\ref{fig:attenuation}.
The data can be described precisely by the following model~\cite{taiuti_measurement_1996}:
\begin{equation}
A(x)=C_0(e^{-x/\lambda} + \alpha e^{(2L-x)/\lambda})
\end{equation}
where L is the length of the bar, $\alpha$ is an empirical factor related to the fraction of photons reflected back from the far end of the bar and adding to the light intensity at the near end.
Based on the fitting result, the light attenuation ratio at various positions can be calculated (see Table ~\ref{tab:summary}) and they are all within our expectation.

\begin{figure}
    \centering
    \includegraphics[width=90mm]{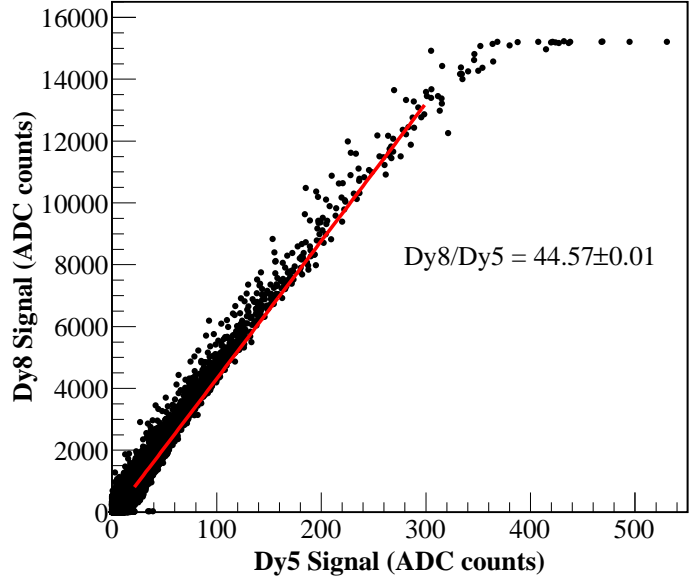}
    \caption{The correlation between the signals recorded by dynode 8 and dynode 5 in cosmic ray test.}
    \label{fig:dy58}
\end{figure}

The gain ratio between Dy8 and DY5 can be extracted from the cosmic ray test by merging data from all tested positions.
The result from the right side of the module is shown in Fig.~\ref{fig:dy58}, and a gain ratio of 44.6 can be derived by using a linear fit. Thus the upper limit of this readout unit can be calculated, and it's about  \SI{1462}{MIPs} for the right side.

The same analysis can be applied to the left side of the prototype module and it has similar results. The measured parameters for both sides are summarized in Table ~\ref{tab:summary}.

\begin{table*}\footnotesize
    \centering
    \caption{Summary of the measured parameters of both ends of the PSD detector module}
    \label{tab:summary}
\begin{tabulary}{190mm}{CCCCCCCCCC}
    \toprule
    &\multicolumn{2}{c}{Electronics(Dy8)}&\multicolumn{2}{c}{MIPs(ADC)}&\multicolumn{1}{c}{}&\multicolumn{2}{c}{Light attenuation ratio}&\multicolumn{2}{l}{$^{40}Ar$(ADC)} \\
    \cmidrule(r){2-3}\cmidrule(r){4-5}\cmidrule(r){7-8}\cmidrule(r){9-10}
    & Calibration(\si{\per\femto\coulomb}) & RMS noise(ADC) & MPV & $\sigma$ & Dy8/Dy5 & FarEnd/Middle & NearEnd/Middle & Mean & $\sigma$ \\
    \midrule
    Left  & 1.129 & 4.5 & 419.6 & 83.1 & 48.1 & 0.76 & 1.86 & 1918.6 & 86.8 \\
    Right  & 1.082 & 5.8 & 395.8 & 65.6 & 44.6 & 0.74 & 1.94 & 1920.6 & 81.3 \\
    \bottomrule
\end{tabulary}
\end{table*}

\subsection{Results from relativistic heavy ion beam test}
\label{sec:beam}
To verify the covering range of the PSD detector module, a heavy ion beam test is carried out at the H8 beam line of CERN SPS, April 2015.
The primary $^{40}{Ar}$ (Z=18) beam of \SI{40}{AGeV\per c} is extracted from SPS and impinging on the middle of PSD's plastic scintillator bar, and the same supply voltage as the cosmic ray test is applied for direct comparison.

\begin{figure}
 \centering
 \includegraphics[width=90mm]{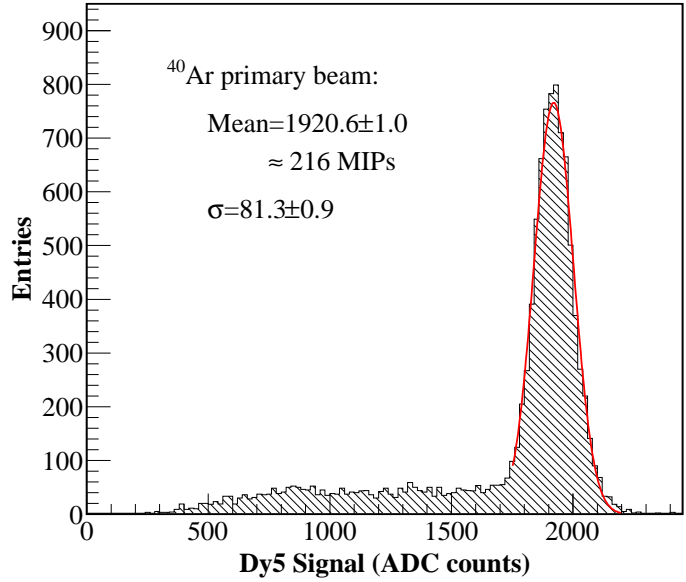}
\caption{Pulse height spectrum of $^{40}Ar$ recorded by dynode 5.}
\label{fig:Ar}
\end{figure}

The signal from Dy8 has found to be already saturated in this measurement and only the data from Dy5 can be used for high charge nuclei analysis.
The ADC spectrum recorded by Dy5 of the right end of the bar is shown in Fig.\ref{fig:Ar}.
The long tail in the left of the primary $^{40}Ar$ peak corresponds to light nuclei which are the products of $^{40}Ar$ interacting with other material in front of PSD detector module.
Using the Gaussian distribution, the mean value of the peak is found to be 1921 ADC counts and the relative resolution is \SI{4.2}{\percent}.
 Combination with the gain ratio between  Dy8 and Dy5 from the former cosmic test, an energy deposition of \SI{216}{MIPs} is obtained. A linear extrapolation based on the simple $Z^2$ dependency (i.e. without considering the quenching effect) gives that the response of Ca (Z=20) shall be about \SI{267}{MIPs}, which is consistent with the calculation result using the parameters from Ref.~\cite{bindi2005performance}. Using the energy fluctuation of \SI{5}{\percent} and taking into account other effects described in Sec.\ref{sec:requirement}, the largest signal of Ca in Dy5 is estimated to be \SI{1167}{MIPs}, which is well within the upper limit of the designed dynamic range. The left side of the module gives similar results, and the data is shown in Table ~\ref{tab:summary}.

\section{Conclusion}
\label{sec:conclusion}
A large dynamic range readout unit with low power consumption has been successfully designed for the PSD detector of DAMPE.
The readout unit could easily cover the dynamic range from 0.1 MIPs to 1400 MIPs, which is need by the PSD for measuring particle from electron to heavy ions with Z up to 20 simutaneously. And the results from tests using cosmic ray and high energy beams match well with the estimation.
However an adjustment of the PMT supplying voltage is required to obtain the optimal performance and make full use of the measurement range. Due to the large individual differences of PMT, it is also suggested that a detailed measurement of the readout PMTs should be performed to reject tubes with extreme characteristics before the massive production and installation.

\section*{Acknowledgement}
\label{sec:acknowledgement}

This work was supported by the Strategic Priority Research Program on Space Science of the Chinese Academy of Science,
Grant No. XDA04040202- 3.
The authors wish to thank all the people from DAMPE collaboration who helped make this work possible.

\section*{References}
\label{sec:reference}
\bibliographystyle{elsarticle-num}

\bibliography{mybib}

\end{document}